\magnification 1200
\baselineskip 18 pt
\topskip .4 in
\leftskip .15in
\rightskip .15in
\vglue .5 in
\vsize 7.0 in

 \def\ea{{\it et al.}}
\def\ie{{\it i.e.}}
\def\h{\hskip 3 pt}
\def\eq#1{ \eqno({#1}) \qquad }
\font\bigggfnt=cmr10 scaled \magstep 3           %
\font\biggfnt=cmr10 scaled \magstep 2            %
\font\bigfnt=cmr10 scaled \magstep 1             %
%
\newbox\Ancha                                     %
\def\gros#1{{\setbox\Ancha=\hbox{$#1$}            %
   \kern-.025em\copy\Ancha\kern-\wd\Ancha         %
   \kern.05em\copy\Ancha\kern-\wd\Ancha           %
   \kern-.025em\raise.0433em\box\Ancha}}          %

\def\Par{\par \vskip .2cm}

\centerline {\bigggfnt Relativistic quantum mechanics} \par\vskip 10 pt
\centerline {\bigggfnt of a Dirac oscillator} \Par

\vskip 18 pt
\centerline {\bigfnt R P Mart\'{\i}nez-y-Romero} \par
\centerline {Departamento de F\'{\i}sica, Facultad de Ciencias} \par
\centerline {Universidad Nacional Aut\'onoma de M\'exico} \par
\centerline {A. P. 21-726 Coyoac\'an, M\'exico 04000 D F, M\'exico} \par
\centerline {E-mail: rodolfo@dirac.fciencias.unam.mx} \Par

\vskip 8 pt
\centerline {\bigfnt H N N\'u\~nez-Y\'epez} \par
\centerline {Departamento de F\'{\i}sica} \par
\centerline {Universidad Aut\'onoma Metropolitana-Iztapalapa} \par 
\centerline {E-mail: nyhn@xanum.uam.mx} \Par

\vskip 8 pt
\centerline {\bigfnt A L Salas-Brito} \par
\centerline{Laboratorio de Sistemas Din\'amicos}\par
\centerline {Departamento de Ciencias B\'asicas} \par
\centerline {Universidad Aut\'onoma Metropolitana-Azcapotzalco} \par 
\centerline {E-mail: asb@data.net.mx} \Par

\vfill
\eject

\centerline{\bigfnt Abstract} \Par 

   The Dirac oscillator is an exactly soluble model recently introduced in the 
context of many particle models in relativistic quantum mechanics. The model
has been also considered as an interaction term for modelling quark 
confinement in quantum chromodynamics. These considerations should be enough 
for demonstrating that the Dirac oscillator can be an excellent example in 
relativistic quantum mechanics. In this paper we offer a solution to the 
problem and discuss some of its properties. We also discuss a physical picture 
for the Dirac oscillator's non-standard interaction, showing how it arises on 
describing the behaviour of a neutral particle carrying an anomalous magnetic 
moment and moving inside an uniformly charged sphere. \Par 

\vskip .1cm

\centerline {\bigfnt Resumen} \Par 

El oscilador de Dirac es un modelo exactamente resoluble que ha sido introducido 
recientemente en el contexto de la mec\'anica cu\'antica relativista de muchos 
cuerpos. El problema ha sido tambi\'en explorado como posible fuente de un 
t\'ermino de interacci\'on para modelar confinamiento en cromodin\'amica 
cu\'antica. Estas consideraciones establecen sin lugar a dudas que el 
oscilador de Dirac puede servir como ejemplo interesante en mec\'anica 
cu\'antica relativista. Este art\'{\i}culo ofrece una soluci\'on al problema y 
la discusi\'on de algunas de sus propiedades. Tambi\'en discutimos una 
im\'agen f\'{\i}sica que se ha introducido para el t\'ermino de interacci\'on 
del oscilador, mostrando como surge de considerar el comportamiento de una 
part\'{\i}cula neutra con momento magn\'etico an\'omalo que se mueve dentro de 
una esfera cargada uniformemente.  

\vskip .3 in

 \noindent Classification numbers: 11.10.Qr; 03.65.Ge; 12.40.Qq. \Par

 \vfill 
 \eject
 \leftskip .25 in
 \rightskip .25 in
 \vskip .5cm

\noindent {\bigfnt 1. Introduction} \par 

Interesting and exactly soluble problems in relativistic quantum mechanics 
(RQM) are scarse. Disregarding the one dimensional examples ---of which the 
book of Greiner (1991) is probably the best source--- the typical problems for 
which the Dirac equation is exactly soluble are (Schweber 1964) the hydrogen 
atom, a particle in an homogeneous magnetic field, a particle in the field of 
an electromagnetic plane wave, and the Morse oscillator. There seem to be no 
new additions to the traditional textbook set in the last few years. Though in 
some situations it would be desirable to discuss examples in RQM (not 
necessarily exactly soluble) more related with contemporary applications, most 
of these are too technical and this usually forbids the discussion. \Par 

Our purpose in this paper is to discuss a problem in RQM, the so-called {\sl 
Dirac oscillator} which, despite its various uses, has an straightforward exact 
solution. Though variants of the model were discussed some years ago (see, for 
example, It\^o \ea\h 1967, Katriel and Adam 1969), the model was recently 
rediscovered in the context of relativistic many body theories (Moshinsky and 
Szczepaniak 1989). The Dirac oscillator has been also studied in connection 
with supersymmetric RQM, with quark confinement models in quantum 
chromodynamics (QCD) and with relativistic conformally invariant problems 
(Moreno and Zentella 1989, Ben\'{\i}tez \ea\h 1990,  Mart\'{\i}nez-y-Romero 
\ea\h 1990, 1991, Mart\'{\i}nez-y-Romero\h and Salas-Brito 1991). On the other 
hand, as we hope to make clear in this work, the problem has interesting 
features which can make it a helpful example in RQM. Besides the Dirac 
oscillator is closely related to the non-relativistic harmonic oscillator, as 
we exhibit in section 2. \Par 

Our paper is organized as follows. In section 2 we introduce the Dirac 
oscillator and discuss its principal properties. In section 3 we offer a 
physical interpretation for the potential of the Dirac oscillator, and cast 
its equation of motion in a manifestly covariant form. A complete solution for 
the problem is given in section 4. We also show that the radial part of the 
components of its spinor eigenfunctions have the same form as the radial 
eigenfunctions of a 3-D non-relativistic harmonic oscillator. Conclusions 
and comments are given in section 5. \Par 

\vskip .25cm

\noindent {\bigfnt 2. The Dirac oscillator} \Par 

\noindent {2.1. Relativistic quantum mechanics} \Par

Before introducing the Dirac oscillator, let us first briefly recall the 
fundamentals of Dirac's RQM (Bjorken and Drell 1964, Greiner 1991). The basic 
equation in the theory is the free particle Dirac equation (for conciseness we 
use units such that $\hbar = c = 1$): 

$$ H_{\rm free} \psi = \left(\gros {\alpha \cdot} {\bf p} + m \beta\right) 
\psi = i {\partial \psi \over \partial t} 
\eq {1} $$ 

\noindent where $H_{\rm free}$ is the free-particle Dirac Hamiltonian, {\bf p} 
= $ - i \nabla$ is the momentum, the $4 \times 4$ Dirac matrices are 

$$ {\gros\alpha}= \pmatrix{{\bf 0}& {\gros \sigma}\cr
   {\gros \sigma}& {\bf 0} },\qquad \hbox{and} \qquad\beta= \pmatrix{ 1& 
0\cr 0& -1}, 
\eq {2} $$

\noindent the $\gros \sigma$ are $2 \times 2$ Pauli matrices and the 1's and 
0's in $\beta$ stand, respectively, for $2 \times 2$ unit and zero matrices. 
If we introduce the gamma matrices $ \gamma^0 = \beta$, ${\gros \gamma} = 
\beta {\gros \alpha}$, then, on multiplying by $\beta$, the Dirac equation can 
be put in the manifestly Lorentz covariant form 

$$ (i \gamma^\mu \partial_\mu - m) \psi = 0. 
\eq {3} $$ 

An electromagnetic interaction is usually introduced in the free particle 
Dirac equation (1) (or (3)) using the standard minimal-coupling prescription 
{\bf p} $\rightarrow$ {\bf p} $- e${\bf A}, and $H \rightarrow H - e \phi$, 
or, in a Lorentz covariant fashion, $p^\mu \rightarrow p^\mu - e A^\mu$, where 
{\bf A} is the vector potential, $\phi$ the scalar electromagnetic potential 
and $A^\mu = (\phi, {\bf A})$. This prescription suffices for writing down the 
Dirac equation for most but not for all the problems, some are not amenable to 
this procedure. For example, for describing the electromagnetic interacion of a 
neutron, the Dirac equation must be written as (Bjorken and Drell 1964) 

$$ (i \gamma^\mu \partial_\mu + {\kappa e \over 4 m_n} \sigma_{\mu \nu} F^{\mu 
\nu} - m_n) \phi = 0, 
\eq {4} $$ 

\noindent where $F^{\mu \nu}$ is the electromagnetic field tensor, $e$ is the 
{\sl magnitude} of the electron charge, $\kappa$ is the anomalous magnetic 
moment and $m_n$ is the mass of the neutron. As we discuss in the next 
section, the coupling term used for the Dirac oscillator can be interpreted as 
an anomalous magnetic interaction. Let us recall that anomalous interactions 
are simply a phenomenological way of describing the effect of residual 
electromagnetic interactions between electromagnetic fields and globally 
neutral composite particles with charged constituents. \Par 

\noindent {2.2. The Dirac oscillator Hamiltonian} \Par

As a simplified model for complex interactions the harmonic oscillator has 
many uses in non-relativistic quantum mechanics,  that it can be exactly 
solved is thus particularly useful. This fact is intimately related with its 
Hamiltonian being quadratic in both the momenta and the spatial coordinates. 
With these considerations in mind and trying to obtain an analogous model in 
RQM, the Dirac equation has been explored in the search of a potential that 
can make it {\sl linear} in both the momenta and the spatial coordinates 
(Moshinsky and Szczepaniak 1989); the resulting problem has been termed the 
Dirac oscillator. To obtain the non-trivial equation of motion for this 
oscillator, linear in both {\bf r} and {\bf p}, we have to perform the 
following non-minimal substitution in the free particle Hamiltonian $H_{\rm 
free}$ 

$$ {\bf p} \rightarrow {\bf p} - i m \omega \beta {\bf r}, 
\eq {5} $$ 

\noindent where $m$ is the mass of the particle described and $\omega$ is the 
oscillator frequency. The Dirac equation for the system is then 

$$ i {\partial \psi \over \partial t} = H \psi = ({\gros \alpha \cdot} ( {\bf 
p} - i m \omega {\bf r} \beta) + m \beta ) \psi. 
\eq {6} $$ 

At this point, we may wonder why the interaction is introduced as we have done 
and not by simply turning on a linearly growing potential in $H_{\rm free}$? 
There are two reasons for this. First, the pure linear term can always be 
gauged away ---to see this, introduce a linearly growing potential to $H_{\rm 
free}$ and take $\Lambda = r^2 / 2$ as a gauge-function for recovering the 
free particle equation. Second, as we show in section 3, the inclusion of the 
$\beta$ matrix is crucial for obtaining a Lorentz covariant interaction. We 
also remark that although the ${\bf p} - i m \omega \beta {\bf r}$ term is 
obviously not hermitian, the complete Hamiltonian remains Hermitian due to the 
presence of the $\gros \alpha$ matrix. The prescription (5) also guarantees 
the C, P, and T invariance properties of the Dirac oscillator (Moreno and 
Zentella 1989). In particular note that the parity invariance of $H$, \ie\h 
the purely spatial invariance under $ {\bf r}'= - {\bf r}$, is fairly 
obvious from equation (6) if you remember that in Dirac's RQM the parity 
operator is 

$$ P= e^{i \varphi} \gamma^0, 
\eq {7} $$

\noindent where $e^{i\varphi}$ is an unobservable phase factor conventionally 
chosen as one of the four values $\pm 1$ or $\pm i$ (Greiner 1991). \Par 

\noindent {2.3. Properties of the Dirac oscillator} \Par

We turn now to stablishing the relationship between the Dirac oscillator and 
the harmonic oscillator. Let us first take the square of the Hamiltonian of 
the Dirac oscillator, in this way, and after a few straightforward 
manipulations, we get 

$$ H^2 = \left({\gros \alpha \cdot} ( {\bf p} - i \omega {\bf r} \beta) + m 
\beta \right)^2 = p^2 + m^2 \omega^2 r^2 + (4 {\bf S} {\gros \cdot} {\bf L} - 
3) m \omega \beta, 
\eq {8} $$ 

\noindent where

$$ {\bf S} = {1 \over 2} {\gros \sigma} 
\eq {9} $$

\noindent is the spin and 

$$ {\bf L} = {\bf r} {\gros \times} {\bf p} 
\eq {10} $$

\noindent is the orbital angular momentum of the oscillating particle 
described. \Par 
 
We can see now that, as equation (8) shows, $H^2$ becomes essentially a 
Klein-Gordon Hamiltonian with harmonic oscillator interaction plus a 
spin-orbit coupling term. In this sense, we may say that the Dirac oscillator 
is something like the ``square root'' of a linear harmonic oscillator. The 
squared Hamiltonian (8) can be used to obtain in a simple way the energy 
eigenvalues of the Dirac oscillator, as we show in section 4. We must remark 
that the squared Hamiltonian (8) is composed only of operators commuting with 
the $\beta$-matrix ---these operators are called {\sl even}. This property of 
$H^2$ implies that a closed form of the Foldy-Wouthuysen (FW) transformation 
can be found for the problem. This useful point is discussed in  
Mart\'{\i}nez-y-Romero\ea\h (1990). \Par 

If we define the total angular momentum of the Dirac oscillator in the usual 
way as ${\bf J} = {\bf L} + {\gros \sigma} / 2$, it is easy to show that the 
system conserves the total angular momentum. To this end, let us show first 
that neither ${\bf L}$ nor ${\gros \sigma}$ are separately conserved: \par 

$$ [{\bf L}, H] = i({\gros \alpha \times} {\bf p}) - m \omega ({\bf r} {\gros 
\times \alpha}) \beta, 
\eq {11} $$ 

\noindent and

$$ [{\gros \sigma} / 2, H] = - i({\gros \alpha \times} {\bf p}) + m \omega 
({\bf r} {\gros \times \alpha}) \beta, 
\eq {12} $$ 

\noindent where $[A,B] \equiv AB - BA$. But, obviously, ${\bf J} = {\bf L} + 
{\gros \sigma} / 2$ does commute with $H$ and, therefore, the total angular 
momentum $\bf J$ is conserved by a Dirac oscillator 

$$ [{\bf J}, H]=0. 
\eq {13} $$
\Par 

\vskip .25cm

\noindent {\bigfnt 3. Physical origin and Lorentz covariance of the interaction 
term.} \Par 

To obtain a physical model for the interaction term in the Hamiltonian of the 
Dirac oscillator we follow here the ideas of Moreno and Zentella (1989), see 
also Ben\'{\i}tez \ea\h (1990). Let us begin with a simple electrostatic 
problem. Consider an uniformly charged dielectric sphere of radius $R$, the 
electric field produced inside the sphere varies as ${\bf E} = - \lambda {\bf 
r}$ ($\lambda$ is a constant), whereas the magnetic field vanishes everywhere 
${\bf B} = {\bf 0}$. We can always consider a very large sphere to safely 
disregard any edge effects. We are going to show how a particle with an 
anomalous magnetic moment coupled to the electromagnetic field produced by the 
sphere leads to the Hamiltonian of the Dirac oscillator. In the process we 
also exhibit the Lorentz covariant properties of the latter. \Par 

In the proper frame of the sphere (let us call this the laboratory frame), we 
may calculate the electromagnetic potential $A^\mu$ associated with its 
static electromagnetic field. Good expressions for this are given by ${\bf A} 
= {\bf 0}$, $\phi = \lambda t r$, that is 

$$ \hat A^\mu_{\rm lab} = \lambda (0, t {\bf r}). 
\eq {14} $$ 

To put this expression in a Lorentz covariant form, we take advantage of the 
gauge invariance of the electromagnetic field. To this end we select the gauge 
function 

$$ \Lambda = - {\lambda \over 4} (t r^2 - {t^3 \over 3}), 
\eq {15} $$

\noindent to obtain the new potential 

$$ A^\mu_{\rm lab} = \hat A^\mu_{\rm lab} - \partial^\mu \Lambda = {\lambda 
\over 4} (t^2 + r^2, 2 t {\bf r}). 
\eq {16} $$ 

\noindent Introducing the unit four-vector $ U^\mu_{\rm lab} = (1, {\bf 0})$ 
---which may be interpreted as the four-velocity of a point particle attached 
to the origin of the laboratory frame (N\'u\~nez-Y\'epez \ea\h 1989)--- we can 
rewrite the 4-potential (16) in the form 

$$ A^\mu = {\lambda \over 4} [2 (U \cdot x) x^\mu - x^2 U^\mu]. 
\eq {17} $$

This expression is manifestly Lorentz covariant, the electromagnetic field 
tensor produced by the sphere can now be calculated as 

$$ F_{\mu \nu} = \lambda (U^\mu x^\nu - U^\nu x^\mu). 
\eq {18} $$

What is the relativistic quantum behaviour of a neutral quantum particle 
moving inside our dielectric sphere? In the laboratory frame only the electric 
field is non-vanishing and, at first sight, it would seem that a neutral 
particle could not interact with this field. But, as we already pointed out in 
section 2, this is not the case if the particle carries an anomalous magnetic 
moment. When we substitute our expression (18) for the electromagnetic field 
in the interaction term appearing in equation (5), we get

$$ { \kappa e \over 4 m} \sigma_{\mu \nu} F^{\mu \nu} = {\kappa e \over 2 m} 
\lambda (i {\gros \alpha \cdot} {\bf r}),
\eq {19} $$

\noindent to obtain an expression for the Hamiltonian in the Dirac equation, 
we only need to multiply equation (19) by the $\beta$ matrix (Greiner 1991). 
What we get in this way is an interaction term which grows linearly with {\bf 
r}. Hence, the Hamiltonian of this problem can also be obtained, as we did in 
section 2 for the Dirac oscillator, by means of a non-minimal coupling 
prescription: 

$$ {\bf p} \rightarrow {\bf p} - {i \over 2 m} e \kappa \lambda {\bf r} \beta. 
\eq {20} $$ 

\noindent We have to choose 

$$ \lambda = {2 m^2 \omega \over e \kappa}, 
\eq {21} $$

\noindent for reproducing the Dirac oscillator equation (6). That is, for 
obtaining a Dirac oscillator the charge density of the sphere, and hence the 
electric field, must be taken as proportional to the Dirac oscillator 
frequency $\omega$. We have thus exhibited that the Hamiltonian of the Dirac 
oscillator may be regarded as describing a neutral particle (a neutron for 
example) with anomalous magnetic moment and interacting with a static radially 
growing electric field. \Par 

\vskip .25cm

\noindent {\bigfnt 4. Eigensolutions for the problem.} \Par 

We now proceed to calculate the complete solution for the Dirac oscillator 
problem. We find that each component of the radial spinorial eigenfunctions 
have the same form as the radial wavefunctions of the 3-D non-relativistic 
oscillator. Hence, the name oscillator applied to the problem may also be
justified at this level. \Par 

As we have seen in section 2, the total angular momentum {\bf J} commutes with 
$H$ and, therefore, the angular momentum is conserved in our system. From 
the rules of sum for angular momentum, we know that $j = l \pm 1 / 2$, where 
$j$ is the total angular momentum and $l$ is the orbital angular momentum 
quantum number, respectively. We also know that in our system parity is a good 
quantum number, so we use it to clasify the energy eigenfunctions. As the 
parity of the spinorial eigenfunctions of radially symmetric problems in 
Dirac's RQM is of the form $(-1)^l$ (Bjorken and Drell 1964), it is 
useful to define 

$$ \epsilon = \cases { +1 & if parity is $(-)^{j+1/2}$, \cr 
                       -1 & if parity is $(-)^{j-1/2}$; } 
\eq {22} $$ 

\noindent in both cases, we have $l = j + \epsilon / 2$. \Par

Since the Dirac oscillator conserves angular momentum, it is convenient 
to split the energy eigenfunction into a radial part and an angular part 
according to ($\pm$ corresponds to the value of $\epsilon$)

$$ \psi ({\bf r}, t) = {1 \over r} \pmatrix { F(r) {\cal Y}^{\pm}_{jml}(\theta, 
\phi) \cr \cr iG(r) {\cal Y}^{\mp}_{jml^\prime}(\theta, \phi) } \exp (-i E t), 
\eq {23} $$ 

\noindent where ${\cal Y}^{\pm}_{jml}$ and ${\cal Y}^{\mp}_{jml^\prime}$ are 
the spinor spherical harmonics (Greiner 1991) 

$$ {\cal Y}^{\pm}_{jml} (\theta, \phi) = \pmatrix {
    \sqrt{l \pm m + {1 \over 2} \over 2 l + 1} Y_{l, m - {1\over2}} (\theta, 
\phi) \cr \cr 
\pm \sqrt{l \mp m + {1 \over 2} \over 2 l + 1} Y_{l, m + {1\over2}} (\theta, 
\phi) }, 
\eq {24} $$ 

\noindent and the $Y_{l,n}$ are standard spherical harmonics. Notice that 
due to the fact that the parity operator $P$ (eq.\ 7) is proportional to 
$\beta (= \gamma^0)$, ${\cal Y}_l$ has to be of opposite parity to ${\cal 
Y}_{l^\prime}$; the only way to accomplish this is by taking $l^\prime = j - 
\epsilon / 2$. \Par 

To obtain the eigenfunctions and the energy spectrum of the Dirac oscillator 
we use $H^2$ (equation (8)). From that equation, we obtain that the 
differential equations governing the behaviour of both components of the wave 
function, are the sum of a harmonic oscillator equation plus a spin-orbit 
coupling term (this is another way to see that the Hamiltonian of the Dirac 
oscillator conserves angular momentum). Notice that for $l = j + \epsilon / 2$ 
we should have 

$$ (j + 1/2) (j + 1/2 + \epsilon) = l (l + 1), 
\eq {25a} $$ 

\noindent whereas, for $l^\prime = j - \epsilon / 2$, we should have 

$$ (j + 1/2) (j + 1/2 - \epsilon) = l^\prime (l^\prime + 1). 
\eq {25b} $$ 

\noindent In this way, we have established that the square of the angular 
momentum satisfies (cf.\ the expression for $\beta$ in equation (2)) 

$$ L^2 = (j + 1/2) (j + 1/2 + \epsilon \beta). 
\eq {26} $$ 

From equation (26), we can get the contribution to the energy spectrum of 
the spin-orbit coupling term. Using the relation 

$$ J^2 + L^2 + 2 {\bf S} {\gros \cdot} {\bf L} + {3 \over 4}, 
\eq {27} $$ 

\noindent and combining equations (26) and (27), we get

$$ {\bf S} {\gros \cdot} {\bf L} = - {1 \over 4} \epsilon (2 j + 1) \beta - {1 
\over 2}, 
\eq {28} $$ 

\noindent which only means that the spin-orbit coupling makes a constant (\ie 
\h independent of {\bf r}) contribution to the energy spectra. Besides, we 
know that for problems with spherical symmetry the square of the momentum is 
related to the square of the angular momentum by the relation (Greiner 1991) 

$$ p^2 = - {d^2 \over dr^2} + {L^2 \over r^2}, 
\eq {29} $$

\noindent valid for radial functions defined like in (23). From this equation 
and (26), we get for the square of the Hamiltonian of the Dirac oscillator 

$$ E^2 - m^2 = - {d^2 \over dr^2} + {(j + {1 \over 2}) (j + {1 \over 2} + 
\epsilon \beta) \over r^2} + m^2 \omega^2 r^2 + m \omega [\epsilon (2 j + 1) - 
\beta]. 
\eq {30} $$ 

This equation shows that the radial solutions of the Dirac oscillator must be 
the same as the solution for the non-relativistic 3-D harmonic oscillator 
(Ben\'{\i}tez \ea\h 1990). This has to be so because, in (30), the term $E^2 - 
m^2$ equals the radial part of a non-relativistic harmonic oscillator 
equation plus the {\sl constant} term $m \omega [\epsilon (2 j + 1) - \beta]$. 
\Par 

Let us now exhibit the eigenfunctions of the problem. For the so-called 
``big'' radial component of the wave function $F(r)$, the energy is positive 
(Greiner 1991) and the orbital angular momentum must be $l = j + \epsilon / 
2$. Using the analogy with the harmonic oscillator, $F(r)$ must be of the form 

$$ F_{n,l}(r) = A (\sqrt{m \omega} r)^{l + 1} \exp (- m \omega r^2 / 2) 
{}_1F_1 (- n, l + 3 / 2, m \omega r^2); 
\eq {31a} $$ 

\noindent For the ``small'' component $G(r)$, the energy is negative and the 
angular momentum is $l^\prime = j - \epsilon / 2$. Consequently, in this case 
the solution must be given by 

$$ G_{n, l^\prime}(r) = A (\sqrt {m \omega} r)^{l^\prime + 1} \exp (- m \omega 
r^2 / 2) {}_1F_1 (- n, l^\prime + 3 / 2, m \omega r^2). 
\eq {31b} $$ 

In these equations, the ${}_1F_1 (a, b, c)$ are confluent hypergeometric 
functions (Davydov 1967, Gradshteyn and Ryshik 1980), $A$ is a normalization 
constant and $n = 0, 1, 2, \dots$ Employing the normalization 

$$ \int \psi^\dagger \psi d^3r = 1, 
\eq {32} $$ 

\noindent the constant $A$ may be easily evaluated, the result is 

$$ A = \left( {m \omega \over \pi} \right)^{1 \over 4} \left[ {n! 2^{n + l - 
\epsilon / 2 + 3 / 2} \over (2 n + 2 L + 1 - 2 \epsilon)!!} \right]^{1 \over 
2} \left[ \left( n + l + 1 - {\epsilon \over 2} \right)^3 + \left( n + l - 
{\epsilon \over 2} \right)^2 \right]^{-1/2}. 
\eq {33} $$ 

To obtain the energy spectrum we proceed as follows: we know that the 
spectrum of a 3-D harmonic oscillator is given by (Davydov 1967) 

$$ E_N = \omega \left( N + {3 \over 2} \right), \qquad N = 0, 1, 2 \dots 
\eq {34} $$ 

\noindent where $N = 2 n + l$ is the principal quantum number. The energy 
spectrum of the Dirac oscillator is thus given by the contribution of the 
harmonic oscillator plus constant terms 

$$ E^2 - m^2 = 2 m E_N + m \omega [ \epsilon (2 j + 1) - \beta]. 
\eq {35} $$ 

Thus the energy spectrum is given by (Ben\'{\i}tez \ea\h 1990): 

$$ E = \{ m \omega [2 (N + 1) + \epsilon (2 j + 1)] + m^2 \}^{1/2}, 
\eq {36a} $$ 

\noindent for the positive energy states, and 

$$ E = - \{ m \omega [2 (N + 2) + \epsilon (2 j + 1)] + m^2 \}^{1/2}, 
\eq {36b} $$ 

\noindent for the negative energy states. It is to be noted that every state 
with quantum numbers $(N \pm s, j \pm s)$ for $\epsilon = - 1$ have the same 
energy than a state with quantum numbers $(N \pm s, j \mp s)$ ($s$ an 
integer). This is a consequence of the dynamical symmetry of the problem 
(Quesne and Moshinsky 1990). The spectrum is shown in figure 1 where this 
property and a characteristic supersymmetryc pattern may be discerned 
(Ben\'{\i}tez \ea\h 1990). It is important to point out here the difference 
between this spectrum and the bound state spectrum of, say, a Dirac hydrogen 
atom. In the latter all the bounded levels are always described by 
admixtures of the big and the small components of the wave function, it would 
not possible then to employ for solving it the explicit separation we used for 
solving the Dirac oscillator problem (Berrondo and McIntosh 1970, Ben\'{\i}tez 
\ea\h 1990). In this sense, we may say that the Dirac oscillator is more akin 
to a free particle than to the Coulomb problem. \Par 

As the above discussion was based in the square of the Hamiltonian of the 
Dirac oscillator and not in $H$ itself, one may wonder if the results 
presented are correct or not. We can show that in fact they are indeed 
correct. We only outline the principal steps. To begin with, recall that 
for systems with spherical symmetry the ${\gros \alpha \cdot} {\bf p}$ term of 
the Dirac equation ---acting on functions defined like in (23)--- can be 
written as 

$$ {\gros \alpha \cdot} {\bf p} = \alpha_r \left[ p_r - {1 \over r} \epsilon 
\left( j + {1 \over 2} \right) \beta \right], 
\eq {37} $$ 

\noindent where 

$$ \alpha_r \equiv {\gros \alpha \cdot} {\bf r}; \qquad p_r \equiv - {i \over 
r} {\partial \over \partial r} r. 
\eq {38} $$ 

After substituting these expressions in the Hamiltonian of the Dirac 
oscillator we obtain, after some manipulations, the system of coupled 
differential equations 

$$ \left\{ - {d \over dr} + {1 \over r} \left[ \epsilon (j + {1 \over 2}) + m 
\omega r^2 \right] \right\} G(r) = (E - m) F(r), 
\eq {39a} $$ 

\noindent and 

$$ \left\{ - {d \over dr} + {1 \over r} \left[ \epsilon (j + {1 \over 2}) + m 
\omega r^2 \right] \right\} F(r) = (E + m) G(r). 
\eq {39b} $$ 

This system of equations can be solved using standard methods (Ben\'{\i}tez 
\ea\h 1990), and the results are exactly the same as those given in Eqs.\ (27) 
and Eqs.\ (32). \Par 

\vskip .5cm

\noindent {\bigfnt 5. Conclusions.} \par 

We have introduced and discussed the properties of a Dirac oscillator. We have 
shown that the Hamiltonian of the Dirac oscillator describes the interaction 
of a neutral particle with an electric field via an anomalous magnetic 
coupling and have found that the energy eigenfunctions are of the same form as 
those of the non-relativistic harmonic oscillator. The reason behind this is 
that the square of the Hamiltonian for the Dirac oscillator becomes 
essentially a harmonic oscillator plus constant terms. We have remarked that 
as the squared Hamiltonian of a Dirac oscillator is composed only of even 
operators then a closed form for the Foldy-Wouthuysen transformation can be 
found. 

We have mentioned also that this system has some remarkable symmetry 
properties. First, as it conserves angular momentum it must be $O(3)$ 
invariant, but in fact that it admits the larger dynamical symmetry algebra 
$SO(3,1) \oplus SO(4)$, producing a more degenerate spectrum than expected 
from $O(3)$ considerations only (Quesne and Moshinsky 1990). This can be seen 
in figure 1. The energy spectrum exhibits also a supersymmetric pattern 
(Ben\'{\i}tez \ea\h 1990,  Mart\'{\i}nez-y-Romero and Salas-Brito 1991). This 
property and its relationship with the existence of an exact FW transformation 
is analysed in 
 Mart\'{\i}nez-y-Romero \ea\h (1990, 1991). The existence of a closed FW 
transformation implies that the positive and negative energy solutions never 
mix, independently of the intensity of the interaction. This characteristic, 
shown by a class of systems besides this one (another example is a particle 
moving in an arbitrary, time independent, magnetic field), has been referred 
to as {\sl the stability of the Dirac sea} ( Mart\'{\i}nez-y-Romero \ea \h 
1990). \Par 

We hope to have made clear that the Dirac oscillator can be profitably used 
when modern examples are needed to illustrate the uses of the Dirac equation. 
It can be exactly solved with no more effort than a non-relativistic harmonic 
oscillator and many of its properties can be calculated and discussed without 
too much elaborate formalism, it can even be solved using purely algebraic 
techniques (Mart\'{\i}nez-y-Romero and Salas-Brito 1991, De Lange 1991a,b). 
Furthermore, its usefulness for modeling confinement in a Lorentz covariant 
way can make its study more motivating for some people. \Par 

\vskip .5cm

\noindent {\bigfnt Acknowledgements} \par 
This work has been partially supported by CONACyT grant 4846-E9406. ALSB 
acknowledges the partial support of Fundaci\'on Ricardo J Zevada.
We want to thank I Campos, A Zentella, J L Jim\'enez and D Moreno for 
their helpful comments on the manuscript. We also want to acknowledge the 
friendly collaboration of F C Bonito, G Pinto, N Humita and, specially, of the 
late F D Micha. \Par 

\vskip .7cm

\noindent {\bigfnt References.} \Par 

\noindent Ben\'{\i}tez J,  Mart\'{\i}nez-y-Romero R P, N\'u\~nez-Y\'epez H N 
and Salas-Brito A L 1990 {\sl Phys.\ Rev.\ Lett.\ } {\bf 64} 1643-1645; {\it 
ibid.\ } {\bf 65} 2085(E) \Par 

\noindent Berrondo M and McIntonsh H V 1970 {\sl J.\ Math.\ Phys.\ } {\bf 11} 
125-141 \Par 

\noindent Bjorken J D and Drell S D 1964 {\it Relativistic Quantum Mechanics} 
(New York: Mc Graw Hill) \Par 

\noindent Davydov A S 1967 in {\it Quantum Mechanics} (Maine: Neo Press) \Par

\noindent De Lange O L 1991a {\sl J.\ Math.\ Phys.\ } {\bf 32} 1296-1300 \Par

\noindent De Lange O L 1991b {\sl J.\ Phys.\  A} {\bf 24} 667-677 \Par

\noindent Gradshteyn I S and Ryshik I M 1980 {\it Tables of Integrals, Series 
and Products} (New York: Academic Press) sections 7.6 and 9.21 \Par 

\noindent Greiner W 1991 {\it Theoretical Physics 3: Relativistic quantum 
mechanics} (Berlin: Springer) \Par

\noindent It\^o D, Mori K and Carrieri E 1967 {\sl Nuovo Cimento} {\bf 51} 
1119-1126 \Par 

\noindent Katriel J and Adam G 1969 {\sl Phys.\ Rev.\ Lett.\ } {\bf 23} 
1310-1312 \Par

\noindent  Mart\'{\i}nez-y-Romero R P, Moreno M and Zentella A 1990 {\sl Mod.\ 
Phys.\ Lett.\ A} {\bf 5} 949-954 \Par 

\noindent  Mart\'{\i}nez-y-Romero R P, Moreno M and Zentella A 1991 {\sl 
Phys.\ Rev.\ D} {\bf 43} 2036-2042 \Par 

\noindent  Mart\'{\i}nez-y-Romero R P and Salas-Brito A L 1991 {\sl J.\ Math.\ 
Phys.\ } {\bf 33} 1831-1836 \Par 

\noindent Moreno M and Zentella A 1989 {\sl J.\ Phys.\ A} {\bf 22} L821-L825 
\Par

\noindent Moshinsky M and Szczepaniak 1989 {\sl J.\ Phys.\ A} {\bf 22} L817-
L819 \Par 

\noindent N\'u\~nez-Y\'epez H N, Salas-Brito A L and Vargas C A 1988 {\sl 
Rev.\ Mex.\ Fis.\ } {\bf 34} 636-644 \Par 

\noindent Quesne C and Moshinsky M 1990 {\sl J.\ Phys.\ A} {\bf 23 } 2263-2274
\Par 

\noindent Schweber S S 1964 {\it An introduction to relativistic quantum field 
theory} (New York: Harper and Row) p 108 \Par

\vfill
\eject

\centerline{\biggfnt Figure Caption}

Figure 1. \par

\noindent The energy spectrum of a Dirac oscillator shown in two different 
ways. For convenience, the numbers given correspond to $(E^2 - m^2) / (m 
\omega)$ instead of $E$. \par 

Notice that states with quantum numbers $(N \pm s, j \pm s)$ for $\epsilon = - 
1$ have the same energy than states with quantum numbers $(N \pm s, j \mp s)$ 
($s$ an integer). This is a consequence of the dynamical symmetry of the 
problem. \par 
 
There is also a typical supersymmetric pattern which is fairly clear in the 
squared spectra shown. The states with positive energy and $\epsilon = + 1$ 
have the same energy than the states with negative energy and $\epsilon = - 1$ 
excepting for the constant term $2 m \omega$. \Par 

\vfill
\eject 
\end